\documentclass[sn-basic,iicol]{sn-jnl}

\usepackage{natbib}

\usepackage{graphicx}%
\usepackage{multirow}%
\usepackage{amsmath,amssymb,amsfonts}%
\usepackage{amsthm}%
\usepackage{mathrsfs}%
\usepackage[title]{appendix}%
\usepackage{xcolor}%
\usepackage{textcomp}%
\usepackage{manyfoot}%
\usepackage{booktabs}%
\usepackage{algorithm}%
\usepackage{algorithmicx}%
\usepackage{algpseudocode}%
\usepackage{listings}%

\newcommand{\Msun}{$\textrm{M}_\odot$} 

\newcommand{\Rsun}{$\textrm{R}_\odot$} 

\newcommand{\Lsun}{$\textrm{L}_\odot$} 

\newcommand\sun{\odot}%
\newcommand{\apj}{Astrophys. J.} 
\newcommand{\apjs}{Astrophys. J. Suppl. Ser.} 
\newcommand{\aap}{Astron. Astrophys.} 
\newcommand{\mnras}{Mon. Not. R. Astron. Soc.} 

\begin{document}

\title[Is there a black hole in the center of the Sun?]{Is there a black hole in the center of the Sun?}


\author*[1]{\fnm{Matthew E.} \sur{Caplan}}\email{mecapl1@ilstu.edu}

\author*[2,3]{\fnm{Earl P.} \sur{Bellinger}}\email{earl.bellinger@yale.edu}

\author[1]{\fnm{Andrew D.} \sur{Santarelli}}
\affil*[1]{\orgdiv{Department of Physics}, \orgname{Illinois State University}, \orgaddress{
\city{Normal}, \postcode{61790}, \state{IL}, \country{USA}}}

\affil[2]{\orgdiv{Department of Astronomy}, \orgname{Yale University}, \orgaddress{
\city{New Haven}, \postcode{06511}, \state{CT}, \country{USA}}}

\affil[3]{\orgname{Max Planck Institute for Astrophysics}, \orgaddress{
\city{Garching}, 
\postcode{85748}, 
\country{Germany}}}

\abstract{There is probably not a black hole in the center of the Sun. Despite this detail, our goal in this work to convince the reader that this question is interesting and that work studying stars with central black holes is well motivated. 
If primordial black holes exist then they may exist in sufficiently large numbers to explain the dark matter in the universe. While primordial black holes may form at almost any mass, the asteroid-mass window between $10^{-16} - 10^{-10} ~ \textrm{M}_\odot$ remains a viable dark matter candidate and these black holes could be captured by stars upon formation. Such a star, partially powered by accretion luminosity from a microscopic black hole in its core, has been called a `Hawking star.' Stellar evolution of Hawking stars is highly nontrivial and requires detailed stellar evolution models, which were developed in our recent work. 
We present here full evolutionary models of solar mass Hawking stars using two accretion schemes: one with a constant radiative efficiency, and one that is new in this work that uses an adaptive radiative efficiency to model the effects of photon trapping.
}


\maketitle

\section{Introduction}\label{sec1}

Asking whether the Sun, or any star, has a central black hole is equivalently a question about dark matter. While the evidence for its existence is now overwhelming, the fundamental nature of dark matter has eluded explanation. It is clear from observations of big bang nucleosynthesis, the cosmic microwave background, large scale structure, cluster lensing, and galaxy rotation curves in spiral galaxies and velocity dispersions in elliptical galaxies, among a long list of other observations, that the majority of the matter in the universe simply does not interact electromagnetically \citep[for a review see][]{young2017survey}. Notably, these observations are independent of one another and span cosmic time, probing the dark matter abundance from the first minutes of the universe to the present, and are all consistent in finding that 85\% of the mass in the universe is invisible and thus is \textit{dark}.

The composition of this invisible matter is unknown and candidates are too numerous to enumerate here, but can be broadly grouped based on their masses. Microscopic theories of dark matter generally consist of new particles beyond the Standard Model \citep[see sec. 27.5 in][]{Zyla:2020zbs}. In contrast macroscopic candidates may be composites of new particles from an entire dark sector, but many candidates only require Standard Model physics such as quark nuggets, strangelets, and most notably primordial black holes \citep{alcock1993possible,PhysRevD.30.272,hawking1971gravitationally}.

Primordial black holes (PBHs), black holes that may have formed in the first instants of the universe, are a popular dark matter candidate. Stochastic density fluctuations in the early universe may have produced overdense regions that collapsed under their own self-gravity, producing large numbers of PBHs \citep{carr2020primordial}. 
While those less massive than about $10^{15}$~g could have evaporated due to Hawking radiation losses by the present age of the universe, those more massive than this limit have barely radiated any mass due to the strong scaling of Hawking radiation luminosity with mass ($L \propto M_{\textrm{BH}}^{-2}$, such that black hole lifetimes scale like $\tau_{\textrm{BH}} \propto M_{\textrm{BH}}^{3}$). Observational constraints on PBHs with monochromatic masses are now quite tight, and restrict PBHs as the entirety of the dark matter to a few mass windows. Notable for this work is the low mass window between $10^{-16} - 10^{-10}$~\Msun, comparable to asteroids \citep{carr2022primordial}.

Despite the claimed constraints, PBHs remain a compelling dark matter candidate for several reasons. 
First, they do not necessarily require new physics beyond the Standard Model to form \citep{Carr_2021}. Second, they have a simple natural production mechanism from the density fluctuations in the early universe. And lastly, there is now suggestive evidence of their existence across large mass ranges \citep[see][for a review]{carr2023observational}. 

If low mass PBHs exist and are numerous then it is possible that stars could capture them \citep{Lehmann_2022}. The suggestion that stars may harbor central PBHs, including the Sun, is well founded in the literature. In the seminal work, \citet{hawking1971gravitationally} estimates that up to $10^{-16}$~\Msun\ of PBHs could have been captured by the Sun, and this idea has long been entertained in the serious literature. For example, before the discovery of neutrino oscillations it was suggested by \citet{1975ApJ...201..489C} that the two-thirds of the Sun's luminosity may be provided by accretion onto a central black hole, rather that fusion, thus explaining the deficit of neutrinos. 

Stars are unlikely to capture PBHs after formation \citep{Montero_Camacho_2019}. A PBH with finite velocity and falling from infinity will be accelerated up above the escape velocity and rapidly transit the star. The transit timescale is small compared to the damping timescales from accretion and dynamical friction, so the PBH is highly unlikely to become gravitationally bound to the star or system. Capture only becomes probable for PBH transits through denser objects such as white dwarfs or neutron stars. Observational consequences of a neutron star capturing a PBH may include fast radio bursts \citep{Abramowicz2018}, for example. Even though capture by main sequence stars is unlikely, it is far more likely that a PBH could become bound to a star upon formation due to the time dependent nature of the gravitational potential of a collapsing cloud. The problem of capture during star formation is hard, but past authors have considered it and shown that it could be quite likely in lower mass halos with slower velocity distributions, especially those found in dwarf galaxies or possibly the early universe \citep{esser2023constraints}. Indeed, arguments are now abundant in the literature showing that if asteroid-mass PBHs exist in large numbers then one should necessarily expect a few to be captured by stars \citep{Capela2013,Capela2014,Montero_Camacho_2019}.

For many readers, intuition from astrophysics will suggest that a star that captures a PBH will be short lived and look nothing like a star during that life. However, we will show that stars with very low mass PBHs could be very long lived with many surviving their entire main sequence phase. Ultimately the evolution is highly sensitive to the accretion physics, which is the subject of the following sections. 

Much of the theory of accretion of stars with central black holes was developed for `quasi-stars,' star-like objects of many thousands of solar masses that may have existed in the early universe powered by accretion onto a central black hole  \citep{Begelman_2008,ball2011structure,Ball_2012}. More recently, these accretion models were adapted for stellar scale main sequence stars with asteroid-mass PBHs in their cores by \cite{Bellinger}. To distinguish traditional quasi-stars from the scenario where a Sun-like star captures a PBH proposed by \citet{hawking1971gravitationally}, \citet{Bellinger} proposed calling these objects `Hawking stars.'  

This work is a primer on Hawking stars and is both a companion and extension of our recent work, \cite{Bellinger}. We begin with an estimate of PBH capture rates in Sec. \ref{sec:rate}.
We present a brief description of our implementation in the stellar evolution code \textsc{Mesa} in Sec. \ref{sec:2}, and some basic theory of black hole accretion in Hawking stars in Sec. \ref{sec:3}. In Sec. \ref{sec:edd} we present stellar evolution models for a range of seed masses using the accretion scheme of \cite{Bellinger}. In Sec. \ref{sec:trap}, we present a new accretion model developed for this work that accounts for photon trapping, along with detailed stellar evolution models. We summarize with some open questions in Sec. \ref{sec:sum}.

\section{Primordial Black Hole Capture Rate}\label{sec:rate}

In this section we briefly estimate the capture rates of PBHs by stars in the Milky Way using the methods of \cite{esser2023constraints} and \cite{esser2023impact}. 
The mean number of asteroid-mass PBHs captured by a star is $\bar{N} = f \eta \, \nu(M)$ where $f$ is the PBH fraction of dark matter, $\eta$ is the `merit factor' of the host galaxy (see below), and $\nu(M)$ is a factor of order unity that depends on the mass of the star. 
The factor $\nu(M)$ is determined numerically from simulations of contraction of star forming clouds by \cite{esser2023constraints}, and takes into account the amount of time required for the PBH to sink into the star. 
They show that $\nu(M)$ is largely independent of PBH mass between $10^{-15}$ \Msun\ and $10^{-13}$~\Msun, but has some weak dependence above this mass. 
It was only determined for stars up to 0.8~\Msun, so future work is required for more massive stars.

The merit factor depends on the dark matter density $\rho_{\rm{DM}}$ and velocity dispersion $\sigma$ of the halo, and dominates the capture probability. If PBHs constitute all the dark matter (${f=1}$) then the merit factor directly gives an order-of-magnitude probability that a star-forming cloud will capture a PBH. The merit factor is
\begin{equation}
    \eta = 
    \left( \frac{\rho_{\rm{DM}}}{ 100 \, \rm{GeV/cm}^3} \right) \left( \frac{7 \, \mathrm{km/s}}{\sqrt{2}\sigma} \right)^3
\end{equation}
\noindent and assumes that only PBHs in the low-velocity tail are captured. For the Milky Way disk, $\eta \approx 10^{-7}$, so it is highly unlikely that the Sun is a Hawking star even if PBHs comprise all the dark matter. However, it does suggest there could be of order $10^4$ Hawking stars in the Galactic disk. In the Gaia Data Release 3 one might expect dozens to hundreds of Hawking stars \citep{GaiaDR3}. In the sections that follow, we will show how these can be searched for.

Perhaps more importantly, \cite{esser2023constraints} shows that most stars in ultra-faint dwarfs like Tucana III (${\eta = 0.51}$) and Triangulum II (${\eta=0.95}$) may have captured PBHs upon formation. This has incredible potential for constraining PBHs in the asteroid mass window as dark matter, but that is only if the evolution and survival times and of Hawking stars are well understood.

We emphasize that more work is needed on PBH captures over a larger range of stellar masses and PBH masses in order to put strong constraints on PBH dark matter using Hawking stars, but these rough order of magnitude scales suffice for motivating this work.

\section{Stellar evolution modeling with central black holes}\label{sec:2}

In the sections that follow we present detailed stellar evolution calculations of Sun-like stars with central black holes. First, we present a brief discussion of our simulation formalism.

The presence of a central black hole has cascading effects on stellar evolution, causing the internal structure of stars with central black holes to slowly diverge from what they otherwise would be in the absence of the black hole. To study this, we will need a model for the accretion physics and feedback between the black hole and the star. 

The current state of the art is \textsc{Mesa}, Modules for Stellar Evolution, a widely used open-source code for simulating stellar evolution in 1D \citep{Paxton2011,Jermyn_2023}. Despite the equations of stellar structure being coupled and nonlinear, a few simple changes in boundary condition along with an accretion scheme is sufficient to model stars with central black holes, and was developed recently by \citet{Bellinger}. 

Consider how the equations of stellar structure are impacted by a central point mass. A central mass changes the inner boundary condition for integration from \mbox{$M(r=0)=0$} to \mbox{$M(r=0) = M_{\textrm{BH}}$}. The black hole of mass $M_{\textrm{BH}}$ is therefore treated as point-like relative to the star, and this is not such a bad assumption. The Schwarzschild radius of a black hole is $R_s = 2 G M_{\textrm{BH}} / c^2$, with constants $G$ and $c$. Stellar massed black holes are order kilometers in size, while an asteroid mass black hole at the evaporation limit is sub-angstrom in size. It should therefore be immediately apparent that the ability of a low mass PBH to grow by accreting matter from its surrounding star is severely limited when compared to a macroscopic stellar black hole. 

Even in the absence of accretion, this one change in boundary changes gravitational gradients throughout the star, and especially in the core. This in turn changes the pressure and density throughout the star. This has further influence on the temperature, and thus nuclear reaction rates. It is immediately apparent that even a simple change such as a small central point mass results in a nonlinear perturbation to stellar evolution. In fact, such a change could also be used to explore how many other families of non-accreting MACHO dark matter could impact stellar evolution if captured at formation, and should be studied in detail in future work.

In addition to this boundary condition change, accreting black holes can radiate strongly. When also including the accretion luminosity, these changes become even more pronounced. 
Matter falling into the black hole at a rate $\dot{M}_\textrm{BH}$ must radiate gravitational potential energy at a luminosity $L$, as in a typical astrophysical environment, at a rate

\begin{equation}\label{eq:Lgen}
L = \frac{\epsilon}{1-\epsilon} \dot{M}_\textrm{BH} c^2    
\end{equation}

\noindent where $\epsilon$ is the radiative efficiency such that the mass-equivalent energy lost from the system is $\epsilon \dot{M}_\textrm{BH}$. This is generally taken to be approximately 0.1 for a non-rotating black hole, but may be as large as 0.30 (or 0.42) for a near maximally rotating (or maximally rotating) Kerr black hole \citep{thorne1974disk,begelman2014accreting}. However, it could also be near 0 if the photons are trapped and unable to diffuse out of the infalling matter sufficiently quickly to escape the black hole \citep{Begelman1979}. For now we will use $\epsilon = 0.08$, which is approximately the energy released falling from infinity to the inner-most stable circular orbit of a non-rotating black hole \citep{Bellinger}. In practice, the assumption of constant radiative efficiency is just a first approximation and will be revisited later in this work (see Sec. \ref{sec:trap}). 

One sees from Eq. \ref{eq:Lgen} that an accretion scheme requires three pieces of physics: an accretion rate $\dot{M}_\textrm{BH}$, an accretion luminosity $L$, and a radiative efficiency $\epsilon$. Expressions for any two of these then set the third, and many schemes for calculating $\dot{M}_\textrm{BH}$ and $L$ are present in the literature, resulting in an adaptive $\epsilon$ and could be simulated in future work \citep{Begelman1979,flammang1982stationary,Flammang1984,Markovic1995}. In the section that follows we will introduce expressions for $L$ or $\dot{M}_\textrm{BH}$ and use them to construct two accretion models.

\section{Black Hole Accretion in the Solar Core}\label{sec:3}

\begin{figure}[ht]%
\centering
\includegraphics[trim={0 0 0 0},clip,width=0.47\textwidth]{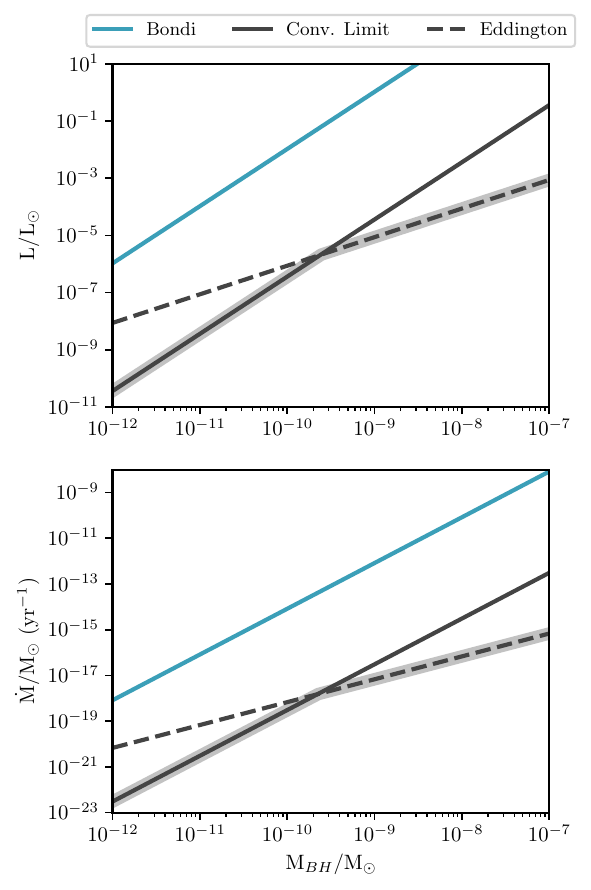}
\caption{(Top) luminosity and (bottom) accretion rate of a low mass black hole accreting from the solar core, assuming a constant radiative efficiency of $\epsilon = 0.08$. For low black hole masses the accretion is limited by the convective luminosity (solid black) while at higher accretion rates the radiation pressure from the accretion luminosity limits the accretion to an Eddington rate (dashed black); they intersect at the Bondi-Eddington transition (BET), $M_{\rm BET} = 2 \times 10^{-10}$ \Msun. The actual evolutionary path is shaded. At no point does this model accrete at the sound speed limited Bondi rate (blue). This may be a useful toy model for a rapidly rotating or magnetized star, and is implemented in the stellar evolution calculations shown in Fig. \ref{fig:kipp1}.  
}\label{fig:scheme1}
\end{figure}

We now introduce our accretion models and show some stellar evolution calculations.  Throughout this section, we will consider the growth of a fiducial $10^{-12}$~\Msun\ PBH accreting material from the solar core. In Fig. \ref{fig:scheme1} we show how the accretion rate and luminosity depend on the black hole mass in our first model, which is discussed in detail below. 

\subsection{Bondi Accretion}

A nominal first estimate for the growth rate would use the Bondi accretion rate. For a black hole embedded in an infinite medium, the Bondi radius is $R_{\textrm{B}}  = 2 G M_{\textrm{BH}} / c_s^2$, with $c_s$ the sound speed of the medium. At this distance from the black hole, the escape velocity is comparable to the sound speed and material from the background begins to freely fall into the black hole \citep{Bondi1952}. This region defines the Bondi sphere containing infalling mass $(8\pi/3) \rho R_{\textrm{B}}^3$ with $\rho$ the density of the medium, and is much less massive than the black hole. This is taken in our stellar evolution models to be a cavity around the black hole from which it accretes, and helps to smooth the numerics at the core \citep{Bellinger}. 

The rate at which matter is accreted is

\begin{equation}\label{eq:dmdtB}
    \frac{dM_{\textrm{BH}}}{dt} = \frac{\pi \rho G^2 M_{\textrm{BH}}^2}{c_s^3} \, .
\end{equation}

\noindent 
This differential equation is exactly solvable, 

\begin{equation}
    M_{\textrm{BH}}(t) = \frac{1}{ M_{\textrm{PBH}}^{-1} - (\pi \rho G^2 / c_s^3)t} \, ,
\end{equation}

\noindent with $M_{\textrm{PBH}}$ denoting the initial mass of the captured PBH. Presently ${\rho = 150~\textrm{g}~\textrm{cm}^{-3}}$ and ${c_s = 5.5\times 10^7~\textrm{cm}~\textrm{s}^{-1}}$ for the solar core. One immediately observes that $M_{\textrm{BH}}(t)$ diverges to infinity at a finite time $\tau_{\rm{B}}$ in the future where

\begin{align}
    \tau_{\rm{B}} &= 1.2~{\rm Myr} 
    \: \left(\frac{\rho(r)}{150~{\rm g/cm}^{-3}}\right)^{-1} \nonumber\\*
    &\times  
    \left(\frac{M_{\rm BH}}{10^{-12}~{\rm M}_{\odot}}\right)    
    \left(\frac{c_{\rm s}}{550~{\rm km/s}} \right)^{3}
\end{align}

\noindent and if correct would immediately suggest searching for stars with central black holes is a fool's errand.

\subsection{Convection-Limited Bondi Accretion}

For an initial $10^{-12}~\rm{M}_\sun$ PBH to grow at the rate given by Eq. \ref{eq:dmdtB} it would also emit $10^{-6}$~\Lsun\ at our fiducial radiative efficiency. This is a luminosity comparable to a very low mass star being emitted from the volume of the Bondi sphere, which is of order 1~mm. The radiation flux produces a steep temperature gradient that drives a convective envelope, stalling the accretion and significantly depressing the accretion rate from this theoretical maximum.

The convective envelope has a theoretical maximum rate at which it can transport luminosity away from the black hole.
In the theory of quasi-stars this convective maximum is

\begin{equation}\label{eq:Lbondicon} 
    L_{B} = 16\pi \eta\, \dfrac{\rho}{c_s \Gamma_1}\, 
    \left(G M_{\textrm{BH}} \right)^2 
    \, .
\end{equation}

\noindent where $\eta$ is the convective efficiency (0.1) and $\Gamma_1$ is the first adiabatic exponent (5/3). If the accretion luminosity exceeds this value then the convective shell will dissipate the energy through shocks and quench the accretion \citep{ball2011structure,Ball_2012}.  

By equating Eqs. \ref{eq:Lgen} and \ref{eq:Lbondicon} 
 one can find a differential equation of the same form as Eq. \ref{eq:dmdtB}. While this equation still diverges at a finite time, it does so at 

\begin{align}\label{eq:tauBcon}
    \tau_{\rm{B}} &= 34 ~{\rm Gyr} 
    \: \left(\frac{\rho(r)}{150~{\rm g/cm}^{-3}}\right)     
        \left(\frac{ (1 - \epsilon) / \epsilon }{11.5} \right)
    \left(\frac{\eta}{0.1} \right)
    \nonumber\\
    &\times  
    \left(\frac{M_{\rm BH}}{10^{-12}~{\rm M}_{\odot}}\right)^{-1}
    \left(\frac{c_{\rm s}}{550~{\rm km/s}} \right)^{-1}
    \left(\frac{\Gamma_{1}}{5/3} \right)^{-1} \, .
\end{align}

\noindent It is now clear that Sun-like Hawking stars, with main sequence lifetimes of 10 Gyr, may live through the entire main sequence phase with minimal changes to the appearance of the star. It is important to note that this is a consequence of the relatively constant core sound speed and density during the main sequence. As a star approaches the giant phase and the core contracts one should expect the rate to drastically increase, necessitating stellar evolution codes to study the post-main sequence. 

\subsection{Eddington Accretion}\label{sec:edd}

\begin{figure*}[ht]%
\centering
\includegraphics[clip,width=0.99\textwidth]{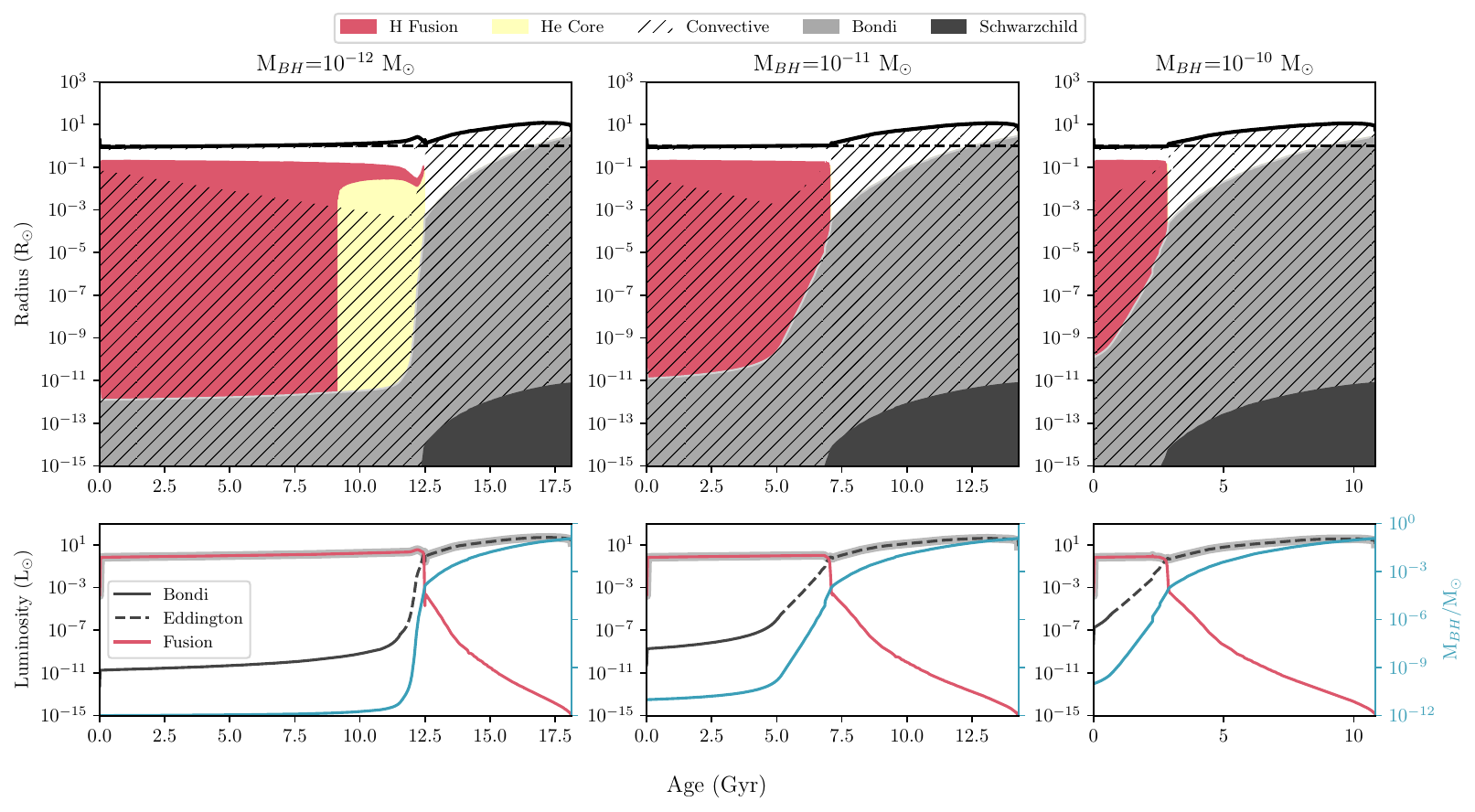}
\caption{Top: Kippenhahn diagrams of stellar evolution models of Sun-like Hawking stars with varying PBH seed masses. Read vertically at a given time one can see the structure of the star from the black hole (dark grey), the Bondi sphere (light grey), and the H fusing core (red). The Bondi sphere and innermost region of the star is convective (slashed) due to the accretion luminosity from the black hole. For the least massive seed simulated, a sizeable He core (yellow) accumulates before the post-main sequence evolution. Bottom: The luminosity due to fusion (red) rapidly plummets as the accretion luminosity (black) comes to dominate the total luminosity of the star (shaded) when the black hole mass is a bit greater than about $10^{-10}$ \Msun. Masses (blue) are given on the right-hand side axis. See also \citet{Bellinger}
}\label{fig:kipp1}
\end{figure*}

We now consider higher mass PBH seeds. While we've seen that $10^{-12}$~\Msun\ PBHs can be hiding in Sun-like main sequence stars, a more massive $10^{-10}$~\Msun\ PBH would seem to destroy the Sun in only 0.34 Gyr from Eq. \ref{eq:tauBcon}. However, the radiation pressure from the accretion luminosity may yet again extend the life of the star. In this simple first model, the accretion rate is not sound speed limited but rather it is radiation pressure limited at the Eddington luminosity, 

\begin{equation}\label{eq:LE}
    L_E =  4 \pi\, \dfrac{c}{\kappa}\, G M_{\textrm{BH}}
    \, ,
\end{equation}

\noindent where $\kappa$ is the opacity of the solar core ($\kappa = 1.5~\textrm{cm}^2~\textrm{g}^{-1}$).  
The accretion rate transitions from a Bondi rate to an Eddington rate approximately when Eqs. \ref{eq:Lbondicon} and \ref{eq:LE} are equal. The black hole mass at the Bondi-Eddington transition (BET) for the solar core is approximately $M_{\mathrm{BET}} \approx 2 \times 10^{-10}$~\Msun\ with an accretion luminosity of $10^{-5}$~\Lsun. More generally, the BET depends on both the metallicity and mass of the star as these set $\kappa$, $c_s$, and $\rho$. We emphasize that this is only a rough first approximation, and the core gas pressure can also play an important role in regulating the accretion rate and should be studied using more detailed accretion schemes like those of \cite{flammang1982stationary,Flammang1984}.

When combining Eq. \ref{eq:LE} and \ref{eq:Lgen}, we obtain a differential equation where $\dot{M}_{\mathrm{BH}} \propto M_{\mathrm{BH}}$, so the black hole grows exponentially as $M_{\mathrm{BH}}(t) = M_\mathrm{PBH} e^{t/\tau_{E}}$ with a growth timescale of

\begin{equation} \label{eq:tauE}
    \tau_{\rm{B}} = 0.15~{\rm Gyr} 
    \: \left(\frac{\kappa}{2~{\rm cm}^{2}~{\rm g}^{-1}}\right)
    \left(\frac{ (1 - \epsilon) / \epsilon }{11.5} \right) \, . 
\end{equation}

\noindent We emphasize that this is not a destruction timescale, but an e-folding timescale. The black hole grows by an order of magnitude in mass on timescales of $\tau_E \ln 10 \approx 0.35 \, \mathrm{Gyr}$, which is long yet again.

Beginning from near the BET ($M_{\textrm{BH}} = 10^{-10}$~\Msun\ and $L_{\textrm{BH}} = 10^{-5}$~\Lsun ), it takes approximately 1.4 Gyr to grow to $10^{-5}$~\Msun. Over the next order of magnitude in mass growth the luminosity of the black hole grows from about $10^{-1}$~\Lsun\ to be in excess of a solar luminosity, thereby modifying the star too drastically for these estimates to be of further use.

In Fig. \ref{fig:kipp1} we show three examples of the evolution of an initially Sun-like Hawking star with a seed PBH in the core calculated with \textsc{Mesa}. For seed masses less than about $10^{-10}$ \Msun, the appearance of the Sun is largely unchanged at present age of the Sun (4.6 Gyr) and the effects of the black hole are confined to the inner core. Eventually, the black hole grows large enough that the accretion luminosity dominates over fusion, causing the star to expand and thus quenching fusion.  At this time the black hole mass is about $10^{-5}$
$\textrm{M}_\odot$, or roughly a Uranus mass. We consider this a useful marker for delimiting the `main sequence' from the `post-main sequence.' 

Once in its post-main sequence phase the Hawking star becomes fully convective and slowly swells over multiple Gyr to about 10 \Rsun\ and about 10 \Lsun\ causing them to appear as a sub-subgiant star, joining a relatively sparse population on the HR diagram. This phase of stellar evolution is qualitatively insensitive to the initial seed mass, and lasts for multiple Gyr in our simulations. 
Simulations with lower constant radiative efficiencies produce similar results, but with shorter post-main sequence phases due to the greater accretion rates (omitted for length).

The transition of a Hawking star from its post-main sequence phase to some other high energy transient is not resolved here, as the \textsc{Mesa} models fail when the Bondi radius reaches the surface of the star ($M_{\textrm{BH}} \approx 10^{-1}$ $\textrm{M}_\odot $). Significantly more work is required to bridge the late stages of evolution in simulations. 

For the lowest mass seed simulated ($10^{-12}$ \Msun) we find that the Hawking star survives for the entire main sequence and even succeeds in forming a He core surrounded by a H-burning shell. At the onset of He core contraction the black hole accretion rate rapidly increases due to the increased core density, sound speed, and conductivity. The resulting increase in accretion luminosity causes the inner convective region to grow outwards, mix the He core with the H-burning shell, and quench fusion. In contrast, those models with more massive seeds reach the post-main sequence before He core formation, so the Hawking star transitions more gradually to its post-main sequence phase with a growth timescale closer to what is predicted in Eq. \ref{eq:tauE}.

\subsection{Photon Trapping and Super-Eddington Accretion}\label{sec:trap}

In this section we introduce the second accretion model that will be considered in this work. 
The Eddington accretion scheme above assumes that photons are not trapped, meaning that photons can diffuse out from the infalling matter in the Bondi sphere faster than the infalling matter can transport the photons inward. However, it is possible the photons are trapped, especially in slowly rotating or weakly magnetized stars. As a consequence, one should expect the radiative efficiency to decrease, especially for higher mass black holes, resulting in rapid accretion at late times.

\begin{figure}[ht]%
\centering
\includegraphics[trim={0 0 0 0},clip,width=0.48\textwidth]{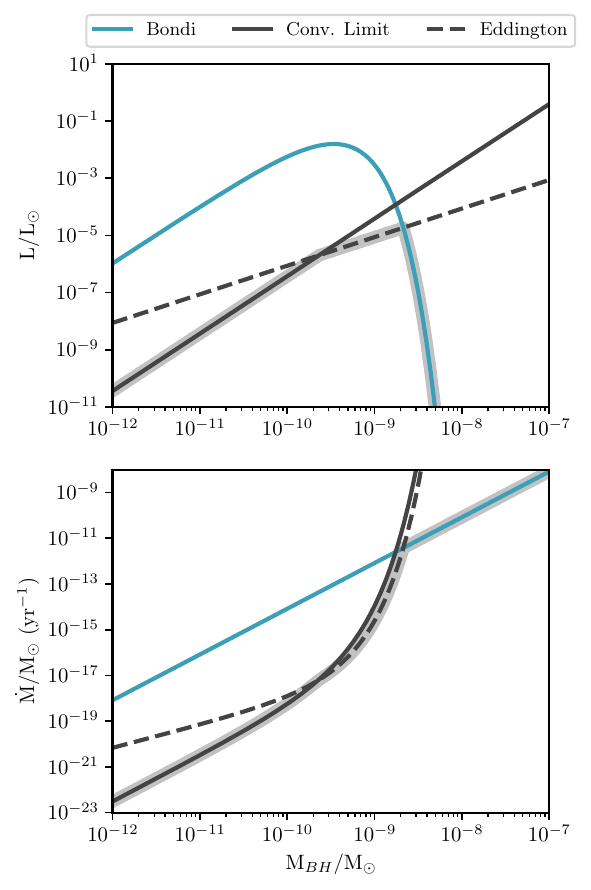}
\caption{(Top) luminosity and (bottom) accretion rate of a low mass black hole accreting from the solar core assuming the adaptive radiative efficiency of Eq. \ref{eq:epsilon2} and fiducial values for the solar core. As the black hole mass grows it more efficiently transports photons inward, reducing the luminosity. The black hole goes `dark' as it rapidly evolves to higher masses by accreting at the Bondi rate given by Eq. \ref{eq:dmdtB}.
}\label{fig:scheme2}
\end{figure}

In Fig. \ref{fig:scheme2} we show a second accretion model using an adaptive $\epsilon$, discussed in detail below. As a brief introduction, this model uses $\epsilon = 0.08$ for low black hole masses but $\epsilon$ rapidly goes to zero above a mass threshold near the BET. To reiterate, there are three variables -- $L$, $\dot{M}_{\mathrm{BH}}$, and $\epsilon$ -- two of which must be set. In both accretion models we set $\epsilon$, leaving one of either $L$ or $\dot{M}_{\mathrm{BH}}$ to be chosen while the other is be determined by Eq. \ref{eq:Lgen}. The convective limit and Eddington limit are imposed by luminosities, while the Bondi limit is imposed by the accretion rate (\textit{i.e.} the maximum rate matter can enter the Bondi sphere). Thus, the convectively limited luminosity and Eddington luminosity are unchanged from Fig. \ref{fig:scheme1}, while the associated accretion rates asymptotically approach infinity with increasing $M_{\mathrm{BH}}$ due to the declining $\epsilon$. Meanwhile, the Bondi accretion rate is unchanged as this is a limit imposed by the sound speed of the ambient accreted material. So, in contrast with the previous accretion model, we see that as the $\dot{M}_{\mathrm{BH}}$ needed to maintain the Eddington luminosity grows exponentially, it eventually reaches a maximum at the accretion rate provided by the Bondi rate.

We now derive our model in detail. 
The mean free path of a photon is $\ell = (\kappa \rho)^{-1} = 4.4 \, \times \, 10^{-4}$~cm for the solar core. Assuming a random walk, such a photon has a diffusion coefficient of $D_\gamma = c \ell / 3$. The diffusion coefficient allows us to calculate a typical photon diffusion timescale $\tau_\gamma$, during which a photon at speed $c$ taking steps of length $\ell$ explores (and escapes from) a sphere of radius $R$ in $\tau_\gamma = R^2 / 6D_\gamma $. Taking $R$ to be the Bondi radius, the photon trapping time is

\begin{align*}
    \tau_\gamma &= 1.1 ~{\rm \mu s}      
    \:     
    \left(\frac{M_{\rm BH}}{2 \times 10^{-10}~{\rm M}_{\odot}}\right)^{2}    
    \left(\frac{c_{\rm s}}{550~{\rm km/s}} \right)^{-4} \nonumber \\
    &\times  
    \left(\frac{\rho}{150~{\rm g~cm}^{-3}}\right) \left(\frac{\kappa}{1.5~{\rm cm}^{2}~{\rm g}^{-1}}\right).
\end{align*}

\noindent To determine if photons are trapped, we compare this to the gravitational free fall time of the infalling matter that the photons are diffusing through. This can be found from Kepler's third law by assuming that a vertical free fall is equivalent to an elliptical orbit contracted to maximum eccentricity. Such an orbit has a semi major axis of $R_{B}/2$, and we take the half-orbit time to be the free fall time,

\begin{align}
    \tau_{\rm{ff}}  = 0.5 ~{\rm \mu s}
    \left( \frac{ M_{\rm BH}}{2 \times 10^{-10}~{\rm M}_{\odot}} \right)
    \left( \frac{c_{\rm s}}{550~{\rm km/s}} \right)^{-3}.
\end{align}

Observe that the photon diffusion timescale and matter free fall timescales are comparable near the BET. Since the diffusion time scales like $M_{\rm BH}^2$ while the free fall time scales like $M_{\rm BH}$, one should expect photons to diffuse out much faster than the free fall time for $M_{\rm BH} \lesssim M_{\rm BET}$ while for $M_{\rm BH} \gtrsim M_{\rm BET}$ the Bondi sphere should be trapping photons, potentially allowing for rapid Bondi-like accretion that is only limited by the supply of infalling matter at the Bondi radius. This would simultaneously increase the accretion rates above the prediction from Eq. \ref{eq:LE} while decreasing the radiative efficiency $\epsilon$, as the matter-energy that is liberated as photons are ultimately consumed by the black hole. 

To incorporate photon trapping in a simple accretion scheme we use an adaptive mass-dependent radiative efficiency rather than fixing $\epsilon=0.08$. For a simple model, one can assume photons must explore and escape the Bondi sphere of radius $R_{B}$ in a free fall time $\tau_{\mathrm{ff}}$, \textit{i.e.} they must diffuse outward faster than the infalling matter can advect them across the event horizon. As photons explore the space they become distributed with respect to the origin, with a radial distribution that could be taken to be a Rayleigh distribution. For small black hole masses, almost all photons escape. At greater black hole masses only those in the tail of the distribution far from the origin escape. The fraction that are trapped can be taken to be the Rayleigh cumulative distribution function $1-\exp(-R_{B}^2/(2\sigma^2))$ where $\sigma$ is a scale parameter with $\sigma^2 = 6 D t_{\mathrm{ff}}$ for a 3D diffusive random walk such that

\begin{align}
    \frac{R_{B}^2}{2\sigma^2}=\frac{G M_{\textrm{BH}} \rho 
  \kappa}{\pi c c_s}.
\end{align}

\noindent Thus, the adaptive radiative efficiency might be (at fiducial values)

\begin{align}\label{eq:epsilon2}
    \epsilon &= 0.08 \exp \left[ -1.15 \left( \frac{M_{\textrm{BH}}}{ 2 \times 10^{-10}~{\rm M}_{\odot}} \right) \right].
\end{align}

\noindent This adaptive radiative efficiency smoothly transitions from near 0.08 at $M\approx 10^{-12}~{\rm M}_{\odot}$ to near 0 at $M\approx 10^{-8}~{\rm M}_{\odot}$.

\begin{figure*}[ht]%
\centering
\includegraphics[trim={0 0 0 0},clip,height=10.25cm]{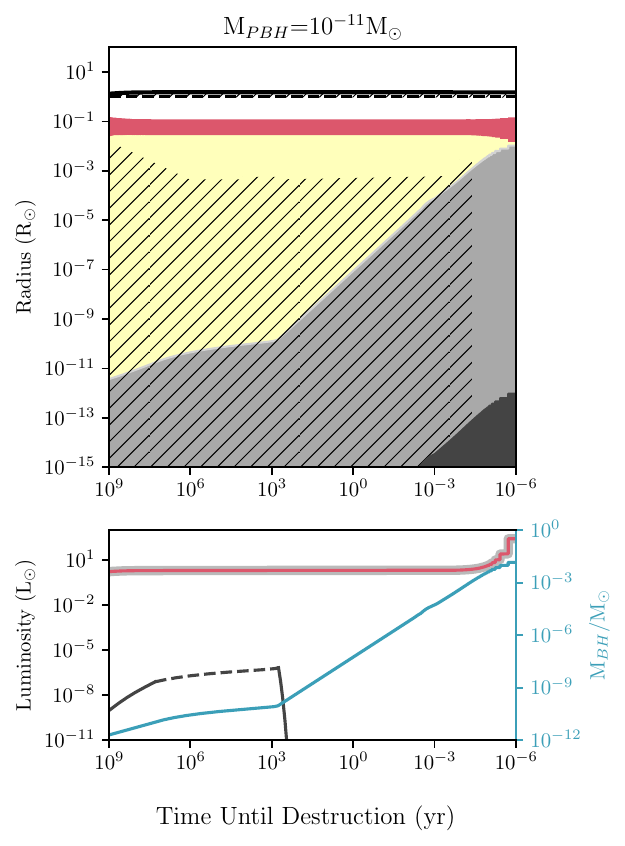}
\includegraphics[trim={0 0 0 0},clip,height=10.25cm]{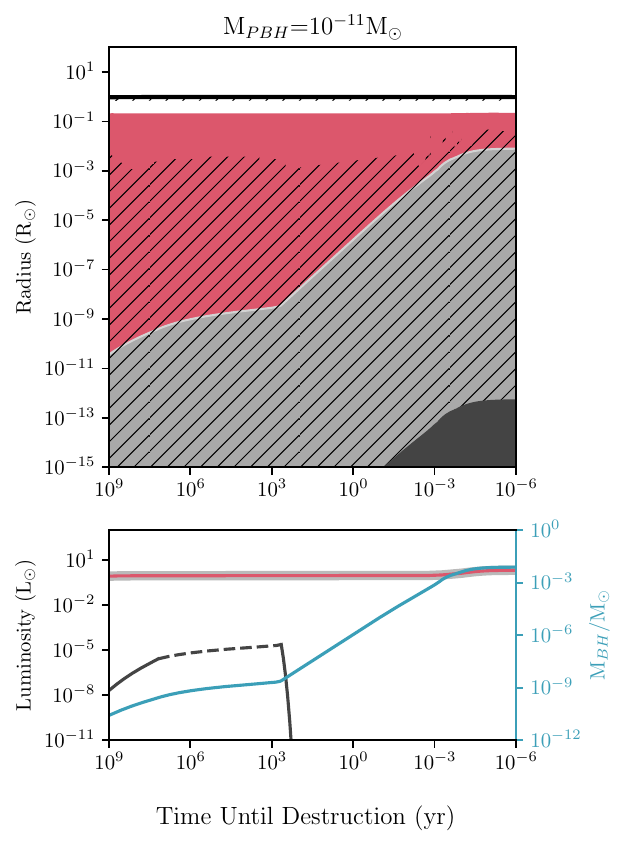}
\caption{ (Top) Kippenhahn diagrams of late stage evolution of a central black hole for a Sun-like Hawking star including photon trapping. Colors and shading is the same as Fig. \ref{fig:scheme1}. Early times, prior to the transition to an Eddington-like accretion, are largely indistinguishable from the models with constant $\epsilon$ in Fig. \ref{fig:scheme1}. The break when about a thousand years remain is associated with the onset of trapping, while the final break with a few hours remaining is associated with a rapid increase in nuclear burning. The time until destruction is taken to be the time until the end of the simulation. (Bottom) The accretion rapidly accelerates as the accretion re-enters the Bondi regime, and the mass grows rapidly to about $10^{-2}$ \Msun{} before the simulation ends for numerical reasons. 
}\label{fig:kipp2}
\end{figure*}

In Fig. \ref{fig:kipp2} we show two stellar evolution models that depict the rapid late time growth of the central black hole with photon trapping for initial $10^{-11}$ \Msun\  and $10^{-12}$ \Msun\ seeds. For numerical reasons both the radiative efficiency and accretion luminosity have a floor of $1.1 \times 10^{-20}$. At early times when $\epsilon \approx 0.08$, the model with an adaptive radiative efficiency is largely indistinguishable from the first model, and for sufficiently low seed masses ($M_{\mathrm{BH}} \lesssim 10^{-12}$ \Msun) the star survives for its entire main sequence and accumulates a large He core. 
However, as a consequence of this new accretion scheme Sun-like Hawking stars do not have an extended post-main sequence evolution.  

The stellar evolution models with an adaptive $\epsilon$ begin to diverge qualitatively from those with a fixed $\epsilon$ near the BET. When the star has about a million years left to live it transitions to a brief Eddington-limited phase, during which the accretion luminosity reaches a maximum at about $10^{-5}$ \Lsun\ before precipitously dropping with the onset of trapping. 
For example, the central convective region in the model with a $10^{-12}$ \Msun\ seed expands out to 1\% of the radius of the star, but does not succeed in mixing the H burning shell with the He core before trapping causes the convective core to shrink inwards. 

The black hole mass at the onset of trapping shows weak sensitivity to the seed mass. The model with the $10^{-12}$ \Msun\ seed has accumulated a large He core, so the higher density causes the onset of photon trapping at a lower black hole mass than the model with a $10^{-11}$ \Msun\ seed. Since higher mass seeds also reach the trapping condition during the H burning phase, the late time evolution is the same as model with a $10^{-11}$ \Msun\ seed.

In the absence of a large accretion luminosity the star does not expand, and the black hole grows by about $10^6$ over about a thousand years with almost no changes to the star's external appearance. In the final hours of the simulation, when the black hole mass approaches 0.01 \Msun, the core temperatures and densities of both models rise by about a half order of magnitude and drive rapid increases the rates of CNO and pp burning. The growth in fusion luminosity can be seen in both models in the bottom panels of Fig. \ref{fig:kipp2} and is thought to play a role in the flattening of $M_{\mathrm{BH}}$ in the final hours.

At this point the accretion timescale and simulation timestep are comparable to the hydrodynamic timescale of the Sun and the simulation stops for numerical reasons. Therefore, even though the final cannibalism of the star proceeds quickly with photon trapping, there may still be interesting transients associated with direct collapse due to the rapid onset of late-time burning that should be explored in future work. While nuclear burning has been largely ignored in past work studying direct collapse of stars with central black holes \citep[e.g.][]{Markovic1995a}, this work finds that it may be important and merits revisiting with detailed stellar evolution models.

\section{Summary}\label{sec:sum}

In this work we have presented some elementary theory of black hole accretion and used it to calculate stellar evolution models of `Hawking stars' -- main sequence stars that may have captured very low mass PBHs in their cores during formation.

The two models presented in this work may be best thought of as extremes, showing how a high radiative efficiency or a low radiative efficiency can drastically influence the evolution toward a long-lived sub-subgiant phase or a rapid direct collapse. It is not immediately clear how suitable either of these models are for describing the real physics of microscopic black hole accretion in a main sequence star in the modern universe. One might reasonably guess that a star could evolve following one of these two qualitative evolutionary tracks depending on the stellar mass, metallicity, rotation, and magnetic field. In this work, we have only explored one fiducial star and a few seed masses using very simplified accretion schemes. Future theory work will be needed to understand accretion in Hawking stars and implement these accretion schemes in stellar evolution codes. Fortunately, a large body of theory on microscopic black hole accretion already exists and could be implemented in \textsc{Mesa} in the near future \citep[e.g.][]{begelman1978accretion,Begelman1979,thorne1981stationary,flammang1982stationary,Flammang1984,Markovic1995,Markovic1995a}.

Observationally speaking, either evolutionary outcome is interesting. While long lived Hawking stars of about a solar mass may spend billions of years as sub-subgiants or red stragglers, they may be distinguishable from traditional stars via asteroseismology due to the fully-convective nature of the star \citep{Bellinger}. Meanwhile, rapid direct collapse provides a mechanism to produce sub-solar mass black holes that are otherwise not predicted by traditional evolutionary channels. As a central black hole does not significantly modify the early phases of evolution, regardless of photon trapping at later times, we suggest that many main sequence stars could be harboring central black holes. Indeed, this work finds that the Sun could be Hawking star with a soft constraint on the PBH seed mass being less than about $10^{-11}$ \Msun.

Many open problems must be addressed to make accurate models of stellar evolution with central black holes; a few are discussed below. In rotating magnetized stars, does the infalling gas form a disk and drive a jet \citep{blandford1982hydromagnetic}? Many stars are rapidly rotating and could have Kerr parameters that are near maximally rotating. Does the black hole get spun up as it accretes, and how does this effect the radiative efficiency \citep{Markovic1995a}? While disks and jets will not be modeled directly in a 1D code like \textsc{Mesa}, their effects could be mocked up by appropriately chosen accretion rates and luminosities. And what is the convective efficiency in the Bondi regime \citep{jermyn2022measures}? What is the radiative efficiency? Do arguments used to derive the radiative efficiency in the continuum limit apply for black holes with sub-angstrom Schwarzschild radii? If not, should the efficiency be greater or lesser, and at what black hole mass do they transition to the canonical efficiency? 
How much mass is lost in winds during the final stages, and what is the final black hole mass? 
These are just a few questions that must be addressed.

Beyond open problems important for the stellar modeling, there are additional questions that must be answered to make the results of stellar evolution calculations astrophysically useful for constraining PBH dark matter. What is the capture rate of PBHs in stars? This will depend strongly on the PBH velocity distribution in star forming halos, which is itself a function a cosmic time. How does the capture rate scale with stellar mass? While this work uses the Sun as a test case, other Hawking star masses should be explored in future work. PBHs may also form in gravitationally bound clusters, which further complicates calculations of their interactions with starforming clouds \citep{belotsky2019clusters,gorton2022effect}. If a PBH becomes bound to a star forming cloud, how long does it take to migrate toward the center? Are there X-ray transients associated with stellar destruction that are distinct from other accreting binary systems, and if so, what are the lifetimes of this transient? Answering these questions is likely beyond the scope of stellar evolution models and simple order of magnitude estimates, and requires detailed calculations.

\bmhead{Acknowledgments}

The authors thank Taeho Ryu, Deepika Bollimpalli, Warrick Ball, Florian K\"uhnel, Selma de Mink, and J{\o}rgen Christensen-Dalsgaard for their contributions to the work this article is based on, and Nicolas Esser for discussion. Financial support for this publication comes from Cottrell Scholar Award \#CS-CSA-2023-139 sponsored by Research Corporation for Science Advancement. This work was supported by a grant from the Simons Foundation (MP-SCMPS-00001470) to MC.

\section*{Declarations}

\begin{itemize}
\item Funding

The funding details are listed in Acknowledgement section.

\item Competing interests

The authors declare no competing interests.





\item Code availability 

Our \textsc{Mesa} implementation is publicly available at {\url{https://github.com/earlbellinger/black-hole-sun}}

\end{itemize}

\end{document}